\pdfoutput=1
\documentclass[10pt]{article}
\usepackage{amssymb,amsmath,latexsym}

\usepackage[sectionbib]{chapterbib}
\usepackage{geometry}

\usepackage[USenglish,english,american]{babel}

\usepackage[utf8]{inputenc}
\usepackage{graphicx}
\usepackage{multirow}
\usepackage{amsmath}
\usepackage{amssymb}
\usepackage{amsthm}
\usepackage{amsfonts}
\usepackage{tipa}
\usepackage{textcomp}
\usepackage{array}
\usepackage{framed}
\usepackage{float}
\usepackage{sidecap}

\usepackage{fix-cm} 
\usepackage{palatino}
\usepackage{caption}
\usepackage{lettrine}

\usepackage[dvipsnames]{color}
\usepackage{listings}
\usepackage{cite}
\usepackage{ctable}
\definecolor{shadecolor}{gray}{0.9}
\usepackage[usenames,dvipsnames]{xcolor}
\usepackage{doi}
\usepackage{url}

\usepackage{authblk}

\title{More than one Author with different Affiliations}
\author[1]{Octavio Miramontes}
\author[2]{Og DeSouza}
\affil[1]{\small{Instituto de Fisica and C3, UNAM, Mexico 04510 DF, Mexico}}
\affil[2]{Department of Entomology, UFV, MG 36570-000, Brazil}

\begin{document}
\title{Social Evolution: New Horizons}
\maketitle
\abstract{
Cooperation is a widespread natural phenomenon yet current evolutionary thinking is dominated by the paradigm of selfish competition. Recent advances in many fronts of Biology and Non-linear Physics are helping to bring cooperation to its proper place. In this contribution, the most important controversies and open research avenues in the field of social evolution are reviewed. It is argued that a novel theory of social evolution must integrate the concepts of the science of Complex Systems with those of the Darwinian tradition. Current gene-centric approaches should be reviewed and complemented with evidence from multilevel phenomena (group selection), the constrains given by the non-linear nature of biological dynamical systems and the emergent nature of dissipative phenomena.
}

\section{Introduction}

Cooperation\footnote{Cooperation here is understood as a nonlinear collective action that results in the benefit of a group.} is everywhere but ecological and evolutionary theories are firmly grounded on competition.  Cooperation is so common and overwhelming in nature that a simple turn of our head will spot it around immediately, appearing in multiple ways and forms. It is so widespread, so much widespread, that it is puzzling why scientists were not willing to easily acknowledge its ubiquitousness and importance. Cooperation and social phenomena are present in humans and in primates and in social insects --the common examples usually given-- but it is also present in unexpected places such as in plants\cite{callaway1997competition, biernaskie2011evidence} or bacteria \cite{griffin2004cooperation, cordero2012ecological} or even as emergent phenomena in artificial societies of robots or other creatures of the cyberspace \cite{langton1997artificial}. Why are we so late in acknowledging this fact? What is the reason of so many years in which biology has lacked a good evolutionary theory of cooperation and social emergence? Charles Darwin was already aware that social behaviour among animals was ``the dirt under the carpet'' for his hypothesis of evolution through natural selection. For him it was so obvious that there was a fundamental and worrisome contradiction in the mere fact of the existence of ant colonies. How can natural selection favour the worker ant that has given up its individuality in the name of the public, anonymous ways of the commune?\cite{cronin1993ant}

It was short after the publication of \emph{On the Origin of the Species} that Herbert Spencer first used the phrase ``survival of the fittest''. The phrase was quickly incorporated into the Darwinian views of biological evolution alongside another masterpiece of ideology uncritically converted into science:``the struggle for survival''. Since these days of the newly-born social Darwinism to the present, little has changed in the mainstream view that social life is a sort of abomination and that the ultimate goal for the living is that egoist, strong and best-fitted individuals be passing their genetic material to future generations, leading to extreme gene-centric views. 

But it needed not to be this way. As lucidly stated by S.J. Gould:\cite{gould2001evolution} ``struggle is often a metaphorical description and need not be viewed as overt combat, guns blazing. Tactics of reproductive success include a variety of nonmartial activities such as [...] better cooperation with partners in raising offspring." In fact, to cooperate rather than defect has been proven the best long-run strategy, even mathematically: game theory predicts that for interactions happening more than once, cooperation is the stable strategy. In order to profit from defection a player has to count on total mindlessness of its partner because an attentive partner will not tolerate recurrent defection. Since memory and cognition are ubiquitous among living beings --at least in rudimentary forms-- mindless partners will not be easily found. The alternative for the compulsive defector is, therefore, to interact with na\"ive partners, but these become experient right after the first interaction! Since the amount of partners is finite, there will come a time when cooperation is the only option. Other mechanisms, besides memory, may stabilize cooperation, even in the face of defectors, such as the system spatial structure \cite{nowak2006five}.

\section{Cooperation at the dawn of life}

\begin{figure}[ht!]
\centering
\includegraphics[width=90mm]{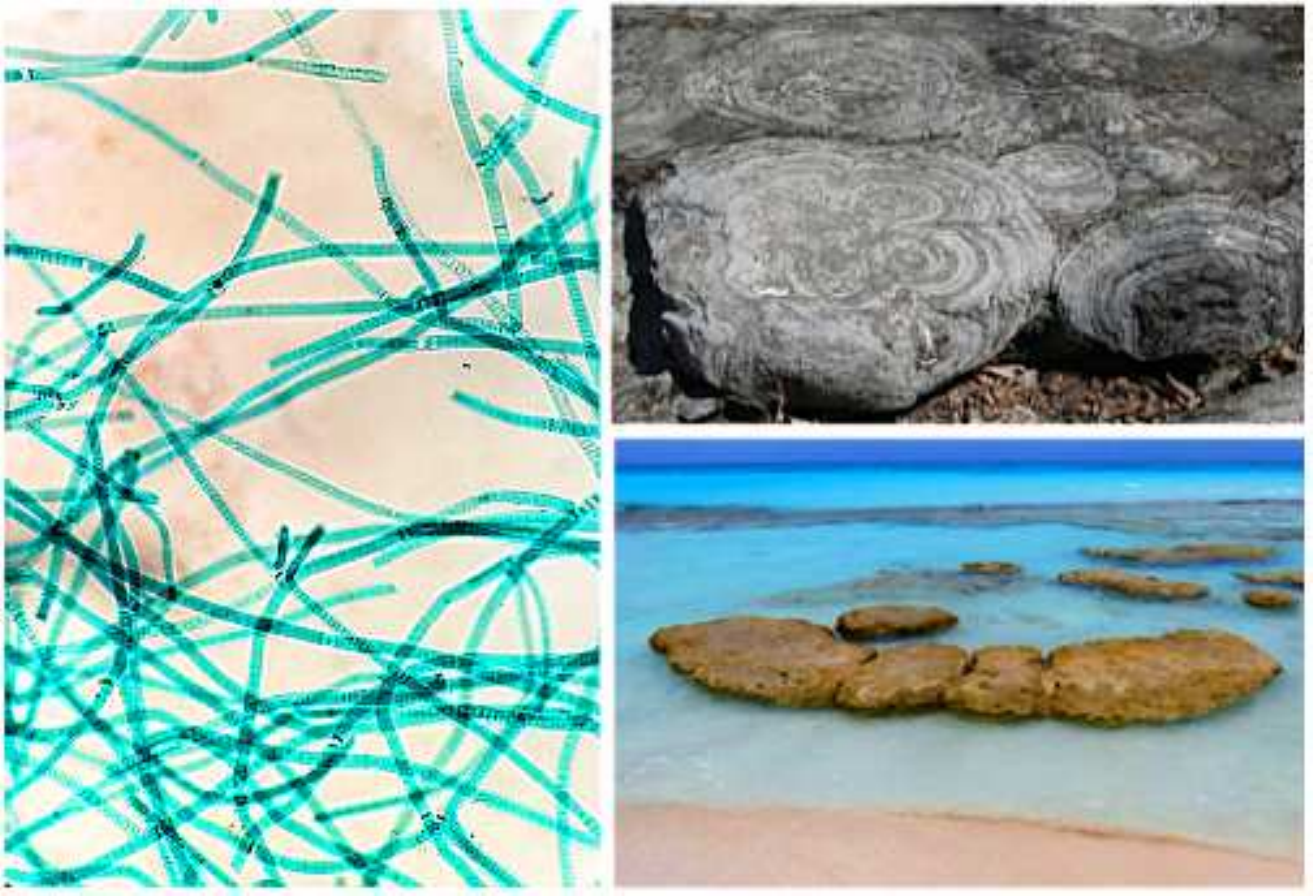}
\caption{{\small 
Fossilized cooperation. One of the most ancient lifeforms on earth is the gregarious filamentous prokaryote known as Cyanobacteria (left). This is a photosynthetic organism that produces oxygen as a by-product of photosynthesis and is worldwide distributed in every ecosystem. Cyanobacteria may form massive aggregations of individuals in shallow waters known as stromatolites. These are layered accretionary structures formed by benthic microbial communities (BMC) where Cyanobacteria are dominant. Stromatolites are the product of dissipative, self-organized systems involving the BMC and its interactions with the environment. Fossilized stromatolites were first described and interpreted circa 1825 as biotic-induced geostructures in the late-Cambrian examples seen near Saratoga Springs, New York, USA (upper-right picture, courtesy of Michael C. Rygel). While declining in number since the Cambrian, stromatolites still can be found in present days as those seen in a Bahamas beach (bottom-right picture, courtesy of Vincent Poirier). Present-day living stromatolites are found in several places around the world, with notable examples at Shark Bay, Australia and at Cuatro Ciénegas, Mexico.   
}}
\label{estromatolites}
\end{figure}

It is unknown how the very first living organisms and their ecosystems on earth looked like. However it is known that the most ancient fossilized organisms were cooperative and social. This is the case of the 3.45 billion years-old Cyanobacteria estromatolites (see Figure \ref{estromatolites}). Cyanobacteria is perhaps the best example of how cooperative behaviour has driven biological evolution. They are suspected to have transformed the initial oxygen-free earth atmosphere into an oxygen-rich one triggering the emergence of aerobic metabolism and ultimately animals. Cyanobacteria are also implicated in the emergence of chloroplasts through a mechanism of endosymbiotic mutualism, a similar cooperative mechanism thought to have originated mitochondria and the eukaryote cell. 

\begin{figure}[ht!]
\centering
\includegraphics[width=90mm]{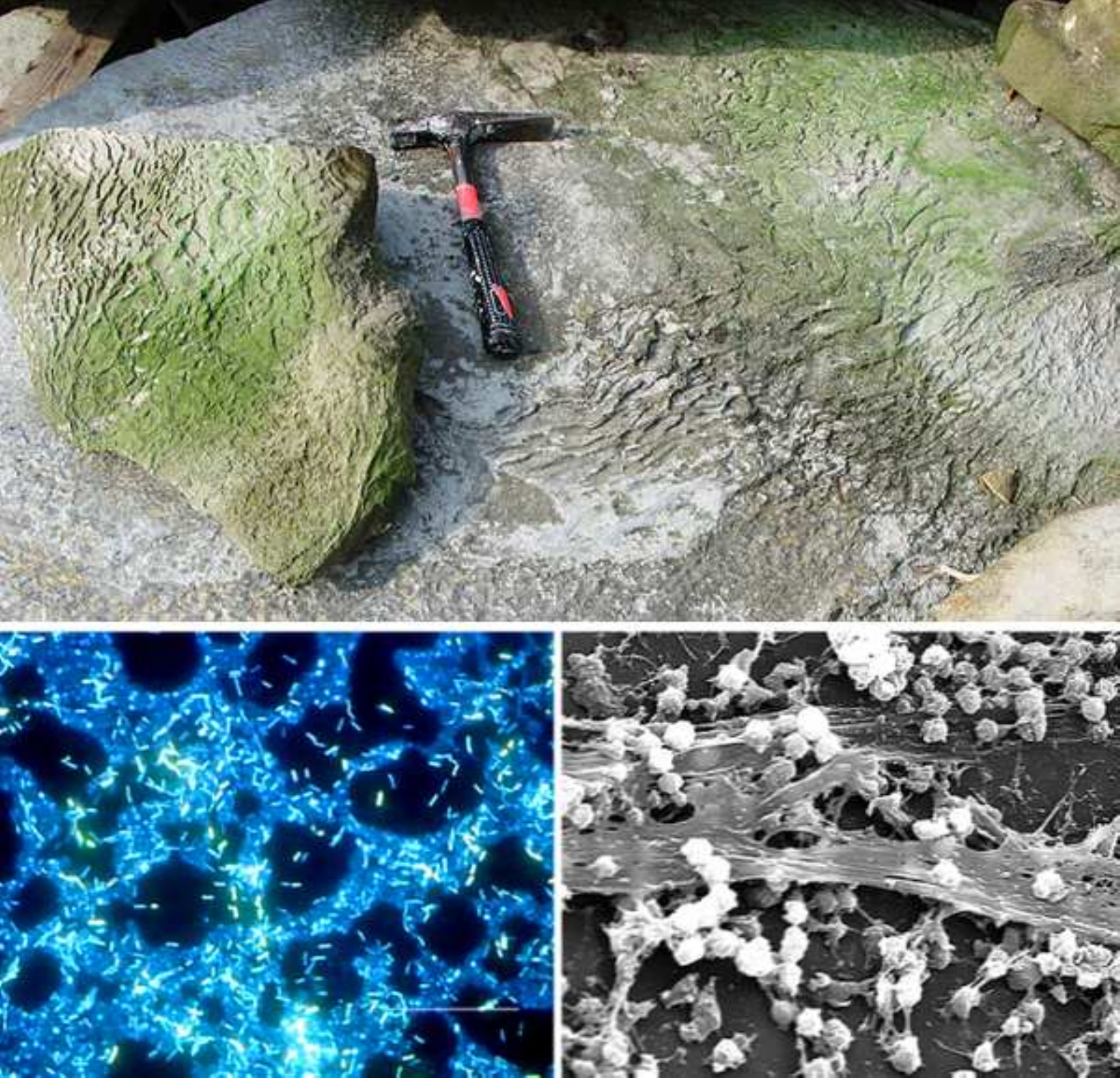}
\caption{{\small 
The oldest evidence of presumably cooperative life on earth are the Microbially Induced Sedimentary Structures (MISS). These are formed by the activity of microbial mats and biofilms (comprised mainly by bacteria). The first MISS described in the literature correspond to those at the Burgsvik Beds in Sweden (upper picture). However the most ancient vestige of biofilm activity in the planet is that at the Dresser Formation, Australia (ca. 3.48 billion-year-old)\cite{noffke2013microbially}. Biofilms and microbial mats are common among present day social bacteria such as those depicted in the lower photographs, corresponding to a polymicrobic biofilm epifluorescence (left) and  an \emph{Staphylococcus aureus} biofilm that has growth at the surface of a medical catheter. Biofilms act as spatio-temporal substrates for the assembly of micro-ecological conditions and social interactions among  prokaryotic multi-species.   
}}
\label{biofilm}
\end{figure}

Stromatolites aside, the most ancient remains of an ecosystem activity currently known have an estimated age of 3.48 billion years-old. These are mineral structures known as Microbially Induced Sedimentary Structures (MISS) and are thought to have been formed by biofilms of single-cell organisms, likely bacteria\cite{noffke2013microbially}. It is worth noting that present day biofilms are well-known paradises for the emergence of complex social interactions and cooperative phenomena among microorganisms (Figure \ref{biofilm}). In fact, it is remarkable that Horizontal Gene Transfer (HGT) was initially discovered in bacteria and that this mechanism of gene transferring is now regarded as a whole new paradigm in evolutionary biology\footnote{This paradigm can be properly named \emph{Post-Darwinian Collective Evolution}. Here we have a mesh of interconnected individual cells that transfer genetic material from cells to cells or that incorporate genetic pieces dispersed in the surrounding environment. This is a socio-environmental scenario where the dynamics of a large interactions network drives evolution without random mutations.  Following what Escalante and Pajares have said in their chapter (this book), the picture that emerges is of the hugest living network ever faced by biologist before. it is the interconnected world of more than  $>10^{30}$ cells (much more than the number of stars in the visible universe) creating dynamically the largest genetic variability collective mechanism ever imagined. }\cite{vetsigian2006collective, goldenfeld2007biology, goldenfeld2010life, buchanan2009collectivist,shapiro2012population, papke2012bacterial}. In order for HGT to work there are at least one essential requirement: that the cells involved in the process come and stay together for a while, socially interacting. This is the reason why biofilms are so important in the early evolution of social life and cooperation. The finding of HGT has triggered many fundamental questions\cite{de2000horizontal, lawrence2002gene}, for example, is the prokaryote diversity and evolution the result of the horizontal exchange of genetic material that allows for the sharing and incorporation of new encoding possibilities, i.e. genetic novelties that are more accessible to selection? or should we stick to the old idea of speciation through random mutations alone \cite{boto2010horizontal}? Let's remember just one thing: HGT is a gene mixing process that occurs between different prokaryotic species and even genera\cite{demaere2013high}.

Closely related prokaryote species show individual genomes that are highly diverse in terms of gene content. As Cordero and Polz reviewed \cite{cordero2014explaining}, much of this variation is associated with social and ecological interactions, which have an important role in the  biology of wild populations of bacteria and archaea. Genetic diversity requires the delineation of populations according to cohesive gene flows (social interactions) and ecological factors, as micro-evolutionary changes arise in response to local selection pressures and population dynamics.

\begin{figure}[ht!]
\centering
\includegraphics[width=90mm]{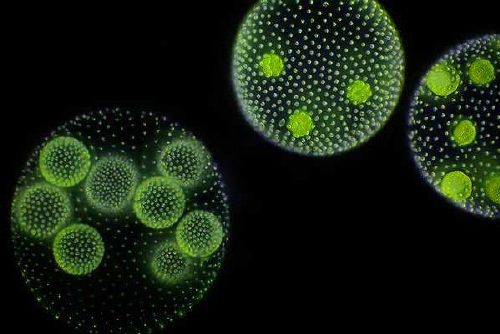}
\caption{{\small 
\emph{Volvox}: a social green evolutionary road. Current evolutionary theory suggests that a  photosynthetic cyanobacterium-like prokaryote was endosymbiotically engulfed by a eukaryotic cell giving rise, eventually, to the entire green plant clade but to green algae in the first place \cite{keeling2010endosymbiotic, de2012diversity}. Later in the evolution of sociality, about 200 million years ago, green algae would assemble into spherical colonies of up to 50,000 cells to form the Chlorophyta. This taxa has an extremely interesting genus called \emph{Volvox} that was among the very first microscopic organisms seen in Antonie van Leeuwenhoek's microscope, circa 1700. Despite all the years that have passed since, there is not a clear idea of how this evolutionary road of cooperation has led to multicelularity, although some recent works point towards environmentally-induced factors as important mechanisms \cite{nedelcu2009environmentally}. A typical \emph{Volvox} colony includes both an asexual cell colony and a sexual one producing microgametes and would also include strong cell differentiation, for example the cells have phototropic eye-spots, which enable the colony to swim towards light. The swimming of the organism occurs in collective coordinated fashion, with many cells being   flagellated. In the picture above, tree \emph{Volvox} individuals. Images like this have given rise to the repeated question of,  how does a cell group evolve into a multicellular individual?\cite{michod2007evolution}. A fundamental question that still remains open; however  emphasis on cooperative mechanisms is increasingly common\cite{queller2009beyond}. 
}}
\label{volvox}
\end{figure}

In the evolutionary history, shortly after the emergence of the prokaryote, single-cell and multi-cell eukaryote emerged but not outside of a cooperative scenario. At a stage somewhere between grouped-individuals and complex multicellular organisms, the Colonial Invertebrates emerged. This is the case of the \emph{siphonophores}, among others, that are highly cooperative forms that integrate multiple individuals, the zooids, into a metazoa that behaves as a single super-organism. Colonial Invertebrates are at the boundary between colonially-grouped and complex multicellular organisms. These are very interesting organisms for the study of social evolution but have been traditionally disregarded since current gene-centric theories of the origin of sociality offers no satisfactory explanations for their evolutionary pathways. \emph{Volvox}, a colonial organism of green-algae is in a similar place regarding its evolutionary origins despite being a model organism for the study of multicelularity evolution (see Figure \ref{volvox}).

Cooperation played a crucial role in the emergence of multicellularity \cite{hochberg2008coevolution}, regarded by John Maynard-Smith and colleges as one of the major transitions in the evolution of life \cite{smith1997major}. It is also interesting to notice that Maynard-Smith regarded the origin of social groups (for example ants, bees, wasps and termites) as another major transition in evolution. He has, however, failed to remark that cooperation is implicated in most of his identified major evolutionary transitions to the point that the opportunity to visualize cooperation as an extremely important force that drives biological evolution, was missed.

Modern post-Darwinian evolutionary theory sees natural selection and randomness as important mechanisms in evolution but argues that these are not the only sources of the extraordinary creativity of nature that we see around\cite{stuart1993origins, goodwin1994leopard, gould2001evolution, miramontes2009evolucion, sacrets2012evolucion}, something else is missing\footnote{See the Chapter by B. Luque \& J. Bascompte and the one by P. Miramontes in this same book.}. Biological evolution did not strictly begin when the first life forms appeared on earth billions of years ago. It is part of a continuum unstopped evolution of matter that started with the big bang and where atoms, molecules and abiotic complex molecules have been built up under the action of non-equilibrium thermodynamics and the physics of complex systems that inspire modern Systems Biology and explains its manifestations: self-organization, emergence, pattern formation, complex networks, dissipative structures, criticality, etc. In what follows we will review where we stand in the construction of post-Darwinian social evolutionary ideas and what we can devise for the future once an integrative view takes into consideration the missing factors of the social evolution of living matter.

\section{Social evolution: the past}
\subsection{The old uncrossed frontier for the ideas on sociality}

In contrast to HGT, Vertical Gene Transfer (VGT) is the mechanism where transmission of genes occurs from the parental generation to offspring via sexual or asexual reproduction. It is under this mechanism that most of the genetic hypothesis for the evolution of sociality and cooperation have emerged in the past, especially as an attempt to describe the emergence of the social life of insects.

As mentioned previously, social insects and their eusociality have always been a challenge for the theory of evolution in Biology. Social colonies are composed of cooperative individuals, most of them subfertile or even sterile, which would not succeed in a world ``red in tooth and claw'' where only the strongest and selfish merciless can prevail. Cooperation and, most notably, reproductive self-denial should have no place in this world. Both traits, however, are too frequent among animals to be simply considered as an insignificant exception. And indeed it is not, as deep analyses of this issue concern scientists, since Darwin himself. To consider recent hot debate on the matter \cite{Abbot.etal.2011,Nowak.etal.2010} this is far from settled, being perhaps the highest mountain pass, a formidable barrier we still need to cross in order to fully understand not simply sociality in insects but the very heart of the theory of evolution.

Examples abound of organisms exhibiting a behaviour in which sterile offspring cohabits with and cooperatively helps their parents to raise fertile offspring, the so-called ``eusociality''. It is found among bees, wasps, ants, termites \cite{Wilson1971Insect}, aphids \cite{Stern.Foster.1997}, thrips \cite{Crespi.1992}, ambrosia beetles \cite{Kent.Simpson.1992}, shrimp \cite{Duffy.etal.2000} and naked mole rats \cite{Sherman.etal.1991}. If this definition is relaxed a bit, allowing senile sterility of parents (as opposed to offspring sterility) and keeping the idea of \emph{group members containing multiple generations and prone to perform altruistic acts as part of their division of labor} \cite[p. 22]{Wilson.2012}, then we may well add even humans to the list of eusocial animals \cite{Foster.Ratnieks.2005}.

\subsection*{The past: puzzles, solutions, and more doubts}

Darwin himself, dedicated a whole chapter to this subject in ``On the Origin of Species'' \cite{Darwin.1859}. Describing the puzzle of the existence of cooperative, sterile individuals in social insects, Darwin comments that they represent \textit{one special difficulty, which at first appeared to me insuperable, and actually fatal to the whole theory.} He circumvented this doubt proposing that 
queens which are able to produce altruistic (cooperative sterile) offspring in addition to ``normal'' fertile ones, would succeed better than those producing only selfish (non-cooperative fertile) offspring, because these latter would not profit from the synergism inherent to cooperative work. In essence, Darwin shifted the problem back to the ``selfish'' scenario, in which the mother queen would be the target of selection. In this sense, the sterile cooperative individuals are a kind of ``extended phenotype'' of the queen, as a fruit is an extension of the mother plant.

Darwin's solution for the evolution of cooperation prevailed for nearly one hundred years, until 1964 when William D. Hamilton advanced an elegant mathematical formalism aimed as an attempt to solve the riddle\cite{Hamilton.1964}. It consisted of the so-called ``kin selection'', which differs from --but does not conflict with-- Darwin's  proposition by establishing that each member of the colony is targeted by selection individually, rather than together with its parents and siblings. Kin selection predicts that individuals cooperate with family members and hence enhance the spread of their genes, indirectly, when their kin reproduce. 
Cooperating within a colony would warrant transmission of genes even for steriles. Such a theoretical construct is sometimes referred to as ``inclusive fitness''.

\begin{framed} 
\small{

\noindent \textbf{\underline{Box 1}. Relatedness in haplo-diploid systems:} 
suppose a fully heterozygous haplodiploid cross:

\begin{center}
    \begin{tabular}{l|r}
        & B\\\hline
      A & AB\\
      a & aB\\\hline
    \end{tabular}
    \end{center}
   
\noindent In Hymenoptera, all offspring produced from this cross are female (males are produced parthenogenetically). Let's take a look at the degree of relatedness between these sisters: 

\begin{center}
\begin{tabular}{l|ll}
sisters & AB & aB\\\hline
AB & 1.0 & 0.5\\
aB & 0.5 & 1.0\\\hline
\end{tabular}
\end{center}
That is, on average, sisters are related to each other by:\\
$1.0+0.5+0.5+1.0=3/4=75\%$%
}
\end{framed}

Haplo-diploidy in Hymenoptera (bees, ants, wasps and sawflies), where males are haploid and females are diploid, was proposed by Hamilton to be the key to the puzzle (see Box 1). A hymenopteran female, by virtue of haplo-diploidy, can share 75\% of its genes with her sisters. Haplo-diploidy, therefore, secures higher levels of kinship between females, which, by abstaining reproduction and helping their mother to raise reproductive sisters, would transfer a load of their own genes which is higher than the load transmitted by their direct reproduction. In a sense, by helping the queen, sterile hymenopteran females almost clone themselves.

Haplo-diploidy, however, is not sufficient to explain the evolution of eusociality: a significant portion of hymenopteran species, while haplo-diploid, are solitary. Maybe more striking, there are many diplo-diploid organisms (having both, males and females, diploid) which are eusocial (Box 2): all the nearly 3 thousand termite species plus aphids, beetles, shrimp, naked mole rats, and humans. Fully diploid organisms do not present kinship asymmetry among siblings, being at most 50\% akin and hence profiting more from their own reproduction than from that of their parents.

It was indeed eusociality in termites --consistently overlooked by texts focusing kin selection-- that always kept alive the challenge, and even more now when the list of eusocial diplo-diploids is frequently updated. Much effort has been made to conciliate termites with kinship selection \cite{Bartz.1980,Bartz.1979,Thorne.1997} but, as noted by Thorne \emph{et al.}\cite{Thorne.etal.2003},  a convincing explanation on why they are eusocial despite their full diploidy is still needed . An important step in this direction was taken by Korb and collaborators\cite{Korb.Heinze.2008}, who presented a broad overview of the ecology of social evolution across large parts of the animal kingdom, including termites \cite{Korb.2008} and other diplo-diploids, thereby expanding the study beyond haplo-diploids. 

\begin{framed}
\small{
\noindent \textbf{\underline{Box 2}. Relatedness in diplo-diploid systems:} in a fully heterozygous diplo-diploid cross we would observe the following offspring:
    \begin{center}
    \begin{tabular}{l|rr}
        & B & b\\\hline
      A & AB & Ab\\
      a & aB & ab \\\hline
    \end{tabular}
    \end{center}

\noindent This will imply in the following degree of relatedness between each of the siblings:

\begin{center}
\begin{tabular}{l|llll}
siblings & AB  & Ab & aB & ab \\\hline
AB & {1.0} & {0.5} & {0.5} & {0.0} \\
Ab & {0.5} & {1.0} & {0.0} & {0.5} \\
aB & {0.5} & {0.0} & {1.0} & {0.5} \\
ab  & {0.0} & {0.5} & {0.5} & {1.0} \\\hline
\end{tabular}
\end{center}
In such case, the average relatedness between siblings is:\\
 $(1*4)+(0.5*8)=4+4=8/16=50\%$%
}
\end{framed}

\subsection*{Contempts}
Meanwhile, it has been argued that the right question has been not posed! In his heavy criticism of the way research has been conducted on kin selection, E.O. Wilson \cite{Wilson.2012} claims to have spotted a philosophical fault in such studies: we have been busy trying to accommodate exceptions to the theory, rather than searching for a better theory that accommodates it all. That is, rather than asking how to conform termites and other diploids to kin selection theory, we should have kept Darwin's first doubt, namely, why are there social --cooperative-- animals in a world apparently ruled by relentless ``struggle for life'' where only the best competitors would survive? Wilson states that we failed to consider multiple competing hypotheses, ignoring well established principles of science philosophy  \cite{Chamberlin.1897}. In Wilson's (2011, pag. 166) own words:

\begin{quotation}
 ``[\ldots] unwarranted faith in the central role of kinship in social evolution has led to the reversal of the usual order in which biological research is conducted. The proven best way in evolutionary biology, as in most of science, is to define a problem arising during empirical research, then select or devise the theory that is needed to solve it. Almost all research in inclusive-fitness theory has been the opposite: hypothesize the key roles of kinship and kin selection, then look for evidence to test that hypothesis.''
\end{quotation}

Stating that Hamilton's rule ``\textit{almost never holds}'', Martin Nowak and collaborators\cite{Nowak.etal.2010} brought recent upheaval to the community of scientists supporting kin selection. It attracted immediate reaction in the form of contentious papers \cite{Abbot.etal.2011,Boomsma.etal.2011,Bourke.2011,Herre.Wcislo.2011,Strassmann.etal.2011}, readily counteracted by supporting ones \cite{Doebeli.2010,Gadagkar.2010,Veelen.etal.2010}. In an attempt to perform neutral analysis of the debate Birch \cite{Birch.2013} identifies ambiguities in Hamilton's defenders and supporters and offers a common vocabulary to help their communication. In short, he states that while kin selection supporters' argument is based on a general form of Hamilton's rule, its opponents construe this rule in a particularly narrow sense. He continues to argue that the current state of deadlock attained by this acrimonious debate will only be broken if both sides agree on common terms\cite{Birch.2013}. 

As an urgent alternative to kin selection as an explanation for the emergence of sociality, Nowak \emph{et al.} \cite{Nowak.etal.2010}, followed by Wilson\cite{Wilson.2012}, proposed that the full theory of eusocial evolution would include the following stages (taken almost \textit{ipsis litteris} from \cite{Nowak.etal.2010}):

\begin{figure}
\centering
\includegraphics[width=70mm]{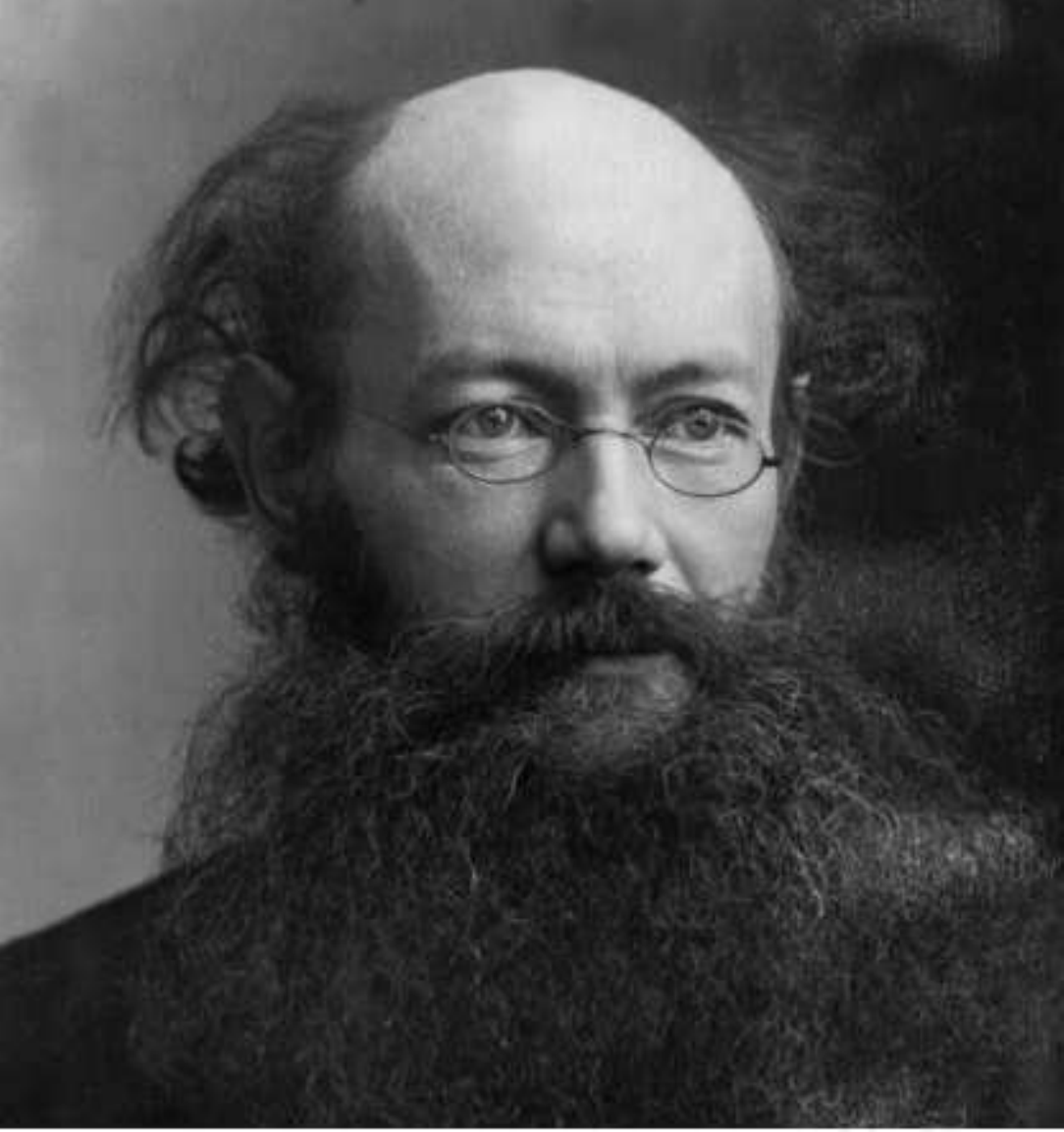}
\caption{{\small 
Forgotten evolutionary thinker. For many years Pyotr Kropotkin (1842–1921) was deliberately ignored by the mainstream of the evolutionary thought until recently when Group Selection was brought forwards again into the modern evolutionary biology school. Kropotkin was a young Russian scientist when he first read ``The Origin of Species'' and felt immediately persuaded by Darwin ideas. Inspired at the age of twenty by the voyages of Alexander von Humboldt, he embarked himself in a long five years exploration of the Siberian lands. While still a Darwinian, Kropotkin had developed his own views on how nature may work. At the time, evolutionary theory developed quickly in England under the conception that the natural world was a brutal place where competition and survival of strongest individuals was the driving force. However, after studying closely flocks of migrating birds, gregarious mammals, fish schools and insect societies, he concluded rightly that competition was almost absent there and that cooperation was indeed common and extended. ``He advocated that natural selection was the driving force that shaped life, but that Darwin's ideas had been perverted and misrepresented by British scientists. Natural selection, Kropotkin argued, led to \emph{mutual aid}, not competition. Natural selection favoured societies in which mutual aid thrived, and individuals in these societies had an innate predisposition to mutual aid because natural selection had favoured such actions''\cite{Dugatkin2011}. Kropotkin moved beyond into considering that the mechanism underlying human cooperation was also the altruistic mutual aid \cite{kropotkin2012mutual}. This observation led him to regard cooperative human societies as self-organized entities that do not need central ruling, being this the essence of anarchism. Due to this, Kropotkin ideas on the evolutionary mechanisms of cooperation were quickly dismissed and regarded as politically unacceptable for the competitive ``free-world'', until today since many current topics on the nature of cooperation were first advanced and investigated by him.  
}}
\label{kropotkin}
\end{figure}

\begin{enumerate}
 \item The formation of groups.
 \item The occurrence of a combination of pre-adaptive traits causing the groups to be tightly formed. Such a  combination would include a valuable and defensive nest, they stress.
 \item The appearance of mutations that prescribe the persistence of the group, most likely by the silencing of dispersal behaviour.
 \item Emergent traits caused by the interaction of group members are shaped through natural selection by environmental forces. 
\item Multilevel selection drives changes in the colony life cycle and social structures, often to elaborate extremes.

\end{enumerate}

In essence, these authors consider groups as an additional unit of selection with selection simultaneously occurring at different levels, e.g., between individuals in the group and between groups. They also remark the importance of spatio-temporal mechanisms that cause individuals to come and stay together. Such ideas are not totally new: they've been long ago hinted by Kropotkin\cite{kropotkin2012mutual} (Figure~\ref{kropotkin}).

\section{Social evolution: the future}
\subsection{Emergent properties of grouping behaviour}
Interestingly, some authors view Nowak and colleagues' proposition as \textit{complementary} rather than \textit{alternative} to kin selection theory. Better stated, they would claim that this ``new'' group-selection theory is in fact a more general proposition of kinship selection \cite{Korb.Heinze.2008a}, despite strong refutation by Wilson \cite{Wilson.2012}. Based on the empirical evidence compiled in the various chapters of their book for a broad range of animals (both vertebrates and invertebrates; full diploids or haplo-diploids), Korb and Heinze\cite{Korb.Heinze.2008} agree with Wilson \cite{Wilson.2012} that the newly re-discovered  group-selection framework is a promising way to investigate the evolution of social phenomena. 

This view would sustain that while in kin selection models relatedness is paramount, the new group-selection models emphasize between-group versus within-group selection, thereby opening an avenue for studies of group level phenomena. Group level phenomena, in which simple repeated interactions between individuals can produce complex adaptive patterns at the level of the group \cite{Sumpter2006principles} are not new in the study of social insects \cite{DeSouza2001,MiramontesDeSouza2008Individual,MiramontesDeSouza1996jtb,Sole.etal.1993}. What is new is the explicit recognition, within the biologists mainstream, that they may hold one important key to help fully understand eusociality. In fact, the awareness of the other (empathy) has been proposed to be one of the traits helping organisms to cross the barrier to sociality and eusociality\cite{de2010age, Wilson.2012}. In humans this would be accomplished by language; in insects by chemical, tactile and visual communication than enhance their ability to interact hence forming cohesive grouping. Interactivity, in fact, seems to be a primary trait underlying grouping in social insects. Depending on the intensity of one-to-one interactions among individuals, complex behavioural patterns can arise at the group level, but these patterns are not hard-wired in these interactions. This should be in fact the next frontier in the studies of eusocial behaviour.

\begin{figure}[ht!]
\centering
\includegraphics[width=120mm]{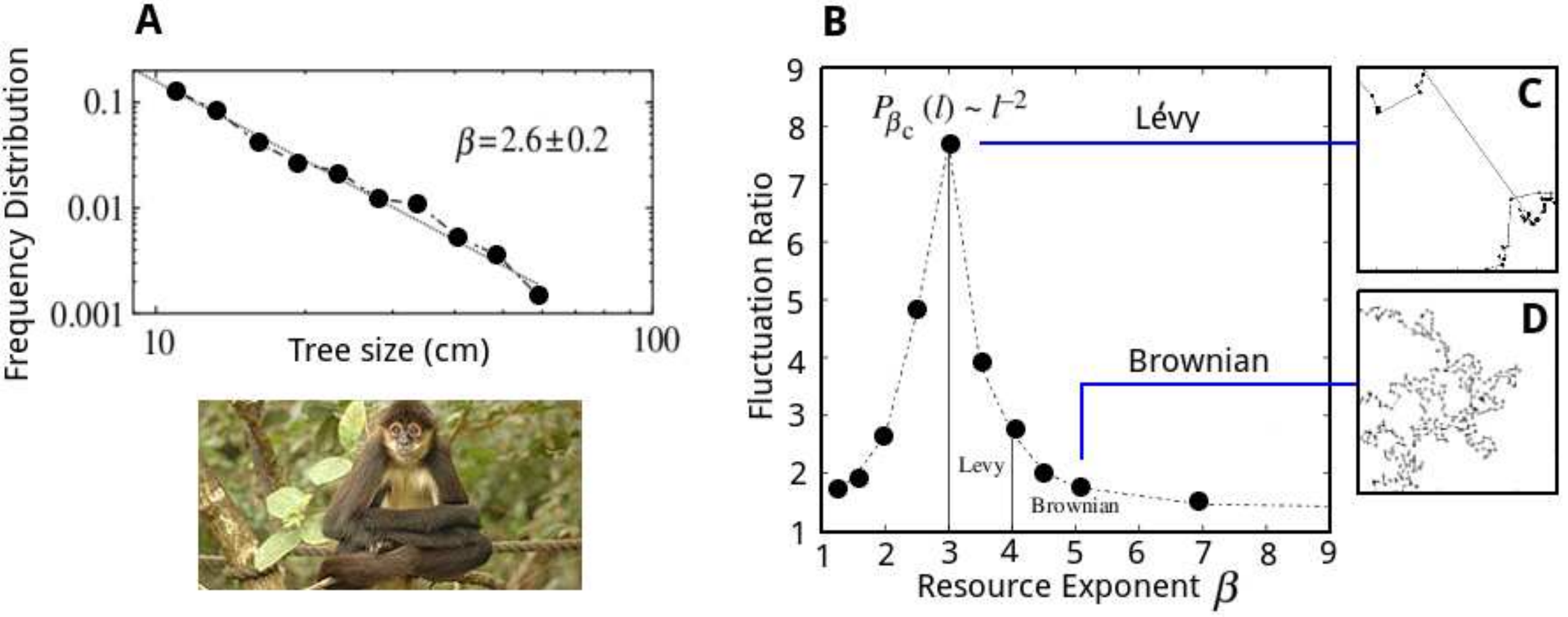}
\caption{{\small 
Emergent anomalous diffusion in social primates. Spider monkeys (\emph{Ateles geoffroyi}) are highly social and forage in groups. (A) In the tropical forest of the Yucatan Peninsula in Mexico, a study of the tree-size frequency distribution showed that this distribution follows a power-law with an scaling exponent value $\beta=2.6$. This fact was used in a foraging model where monkeys move to fruiting trees following a simple optimization rule of ``move to the richest but nearest'' tree (B). The model predicts that the monkey mobility is emergent as anomalous diffusion (Lévy-like) for $\beta$ values close to the observed true value $\beta=2.6$ (C) and is normal diffusive (Brownian) for other values of $\beta$ (D). The emergent nature of the anomalous diffusion is due to nothing else but to the forager-environment interaction\cite{boyer2006scale, boyer2009levy}. It has been also noted that Lévy displacement distribution may bring optimal efficiency to the foraging process\cite{ramos2004levy}.        
}}
\label{monos}
\end{figure}

\begin{figure}[ht!]
\centering
\includegraphics[width=120mm]{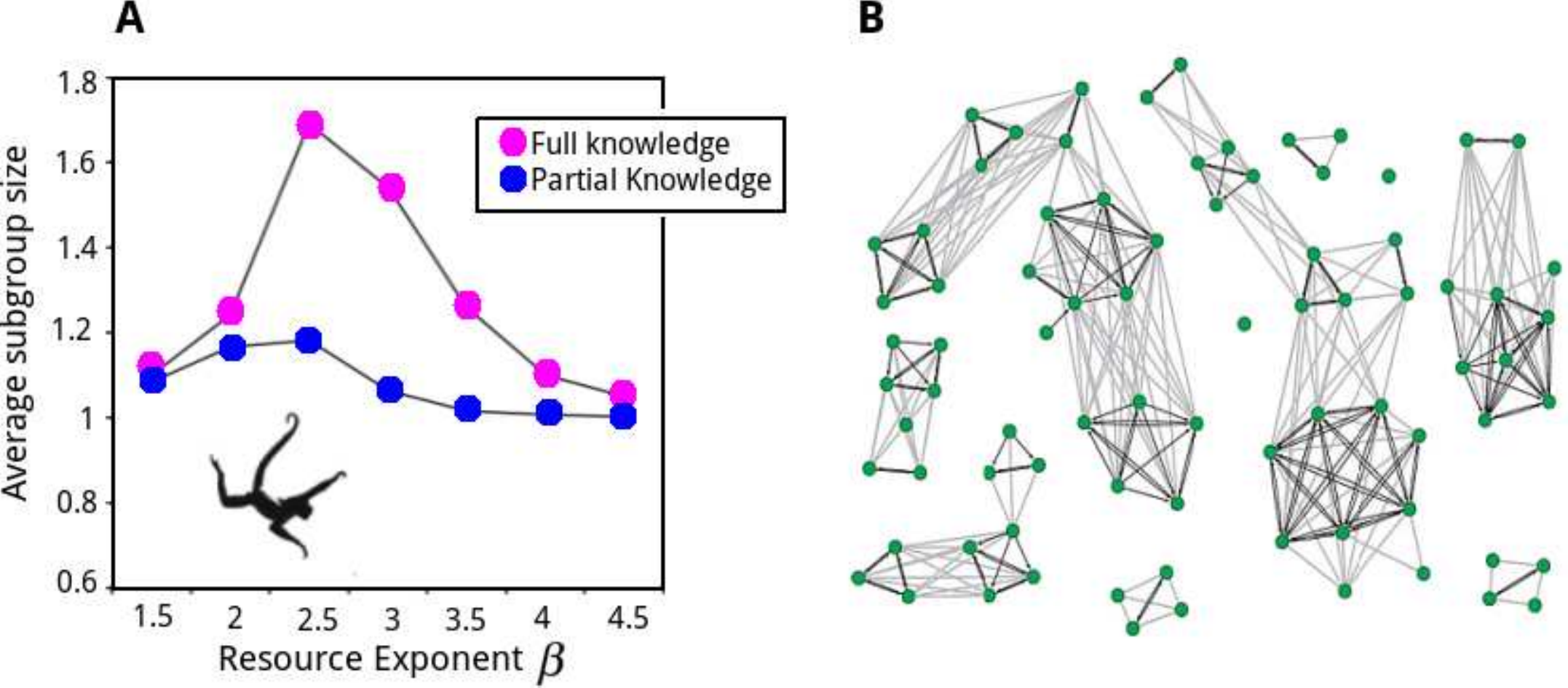}
\caption{{\small 
Spider monkeys emergent social networks. (A) The same foraging model described in Figure~\ref{monos} is now used in a multi-agent environment\cite{ramos2006complex}. It has been found that monkeys foraging in the Lévy anomalous diffusion regime (resource exponent $\beta=2.6$) spontaneously form social ties with the largest average group size. These ties transform into emergent social networks (B). These complex networks would only exist because the monkey mobility is Lévy distributed\cite{ramos2006complex}. The transition to an anomalous diffusion in this foraging model may be interpreted as an order-disorder phase transition\cite{sole2011}.    
}}
\label{monos-net}
\end{figure}

\subsection{Non-randomness and interaction dynamics}

Random mutations are at the centre of current evolutionary paradigm. While it is true that bacteria, for example, adapt and develop resistance to almost every antibiotic that is developed, not a single new species has been observed to arise after decades (hence, thousands or millions of generations) of laboratory experiments in which bacteria are exposed to mutagenic forces. Most of mutations seems to be neutral and do not provoke major inheritable changes that could trigger observable speciation. In fact long-term studies on observable bacterial adaptation suggest that fitness changes in bacteria may occur primarily by the accumulation of neutral mutations\cite{leiby2014metabolic}.

Are random mutations a real mechanism for genetic variation and evolutionary change or are they part of a limited working hypothesis that must me revisited and complemented with, for example, mutationless evolution\cite{huang2012tumor,pisco2013non} or evolution by means of horizontal recombination mechanisms\cite{fraser2007recombination}? An illustrative example regarding random mechanism in theoretical ecology is useful at this point. For years it was thought that random climatic variations were responsible for driving population dynamics. However after the pioneering work of Robert May and others\cite{may1976simple}, it became clear that variations in population numbers may be due to the intrinsic changes of the ecosystems and the non-linear universe of interactions on it. These erratic fluctuations are not random but chaotic and the difference between these concepts is not trivial. One is the outcome of stochastic casino-like events while the second is the outcome of a dynamical complex system with determinism embedded. Is it time to start looking for signs of deterministic dynamical systems as sources of genetic variation?\footnote{See the Chapter by Pedro Miramontes in this same book.}   

\subsection{Mobility: come and stay together}

Another common widespread idea in biology is that individual social and ecological interactions follow essentially random patterns. Take for example mobility and dispersal. Since the 70s of the last century, it became theoretically obvious that a simplified agent would explore its surrounding space efficiently when moving in a fractal pattern that result in what is known as anomalous diffusion (Figure \ref{monos}). Such an efficient pattern would result in increased encounter with prey or con-specifics. This would ultimately lead to increased reproduction rates (genetic diversity), either by obvious positive effects of increased food intake or by the less evident consequences of increased social interaction rates. Because such interactions translate in maximized information flow and processing, they promote efficient cooperation and hence social facilitation (Figure \ref{monos-net}) leading to maximized survival under strong stresses (Miramontes and DeSouza 1996, Rosengaus et al. 1998, Desouza et al. 2001) and hence extending opportunities for reproduction.

\begin{figure}[ht!]
\centering
\includegraphics[width=90mm]{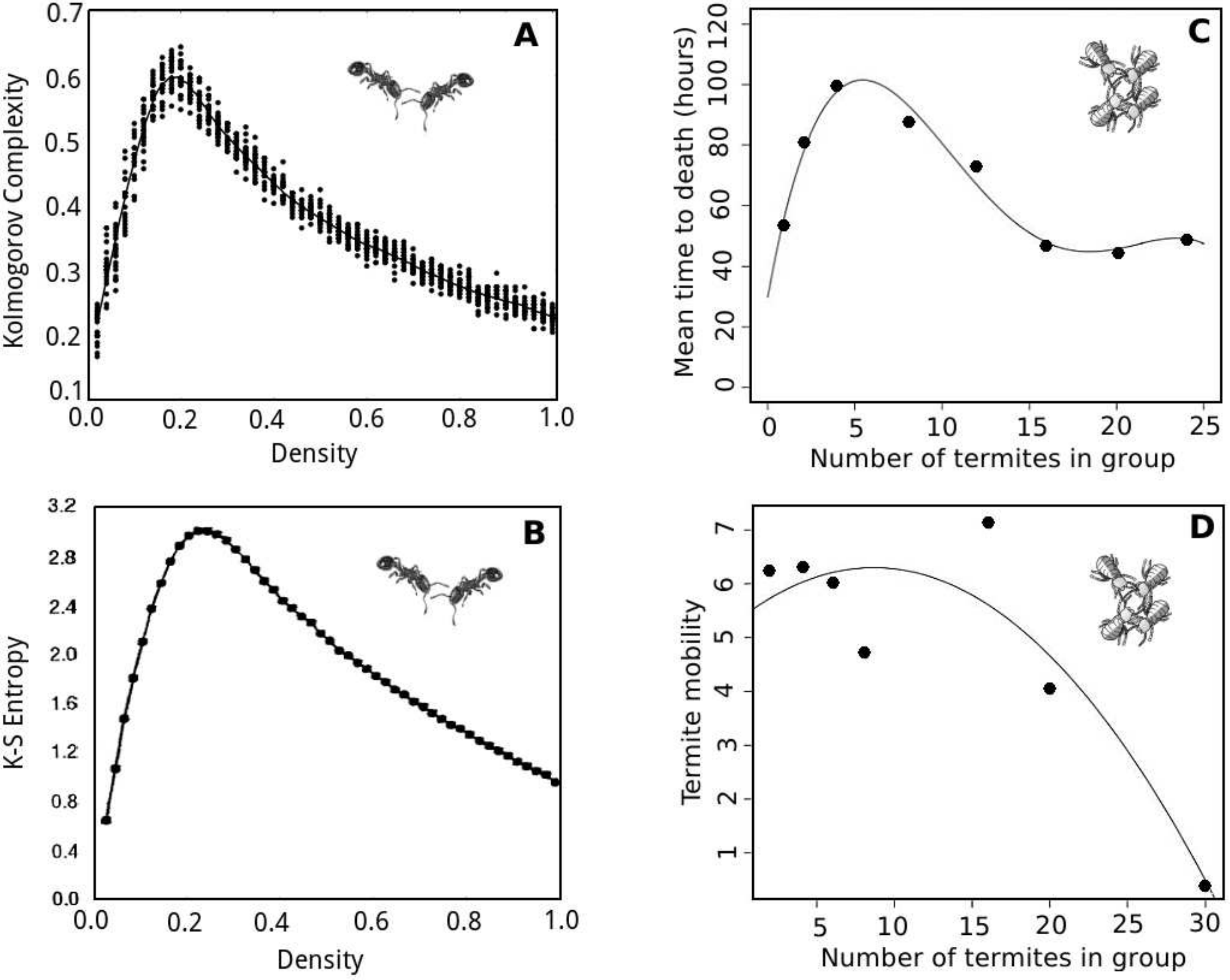}
\caption{{\small 
Social phase transitions in ants and termites. Social insects are good examples of self-organized societies exhibiting a range of group-size complex behaviours involving the criticality properties normally associated with order-disorder phase transitions. Ants of the genus \emph{Lepthotorax} are known for displaying periodic pulses of activity in the colony when measured as movement inside the nest. However the individuals present low-dimensional chaotic movement activity. Then, as the density on the nest increases there is an order-disorder (edge of chaos) phase transition that can be explored when modelled with agent-based formalisms. In such ant models the phase transition occurs at a density that maximize the information transfer and the diversity of observable behaviours of the individuals as measured by (A) Kolmogorov complexity and (B) by a KS-Entropy\cite{miramontes1995order, miramontes2001neural}. Experimental procedures reveal the presence of phase transitions in termite social behaviour. (C) Social-facilitated survival in size-dependent groups of termites show a peak at low densities (C) that is also in agreement with individual mobility (D)\cite{desouza2004non}.    
}}
\label{insects}
\end{figure}

There is growing evidence that biological organisms perform anomalous diffusion in their mobility patterns in the form of Lévy flights (scale-free probability distributions in the lengths of travelled distances). When efficient social interactions occur in the context of density-dependent ecosystems then another interesting phenomena emerges: the so called order-disorder phase transitions that are becoming a new paradigm in evolution\cite{stuart1993origins, goodwin1994leopard, torres2012criticality}.   

Cyanobacteria, as said before, are intriguing social organisms that have been protagonists of important evolutionary changes in the history of life. Despite its apparent simplicity, they are known to have very complex patterns of non-random mobility, cell-to-cell interactions and communication\cite{galante2012modeling}. Cyanobacteria do form pairwise ensembles of interaction and mobility. It would not be surprising at all that their mobility patterns are anomalous diffusion and so their social engagements may respond to optimized encounter rates. It will be also very interesting to know how and when these patterns have emerged in the evolution of these ancient organisms. Have these mechanisms been also present in the mobility and dispersal of interacting proto-cells or self-replicating biomolecules?

\begin{figure}[ht!]
\centering
\includegraphics[width=140mm]{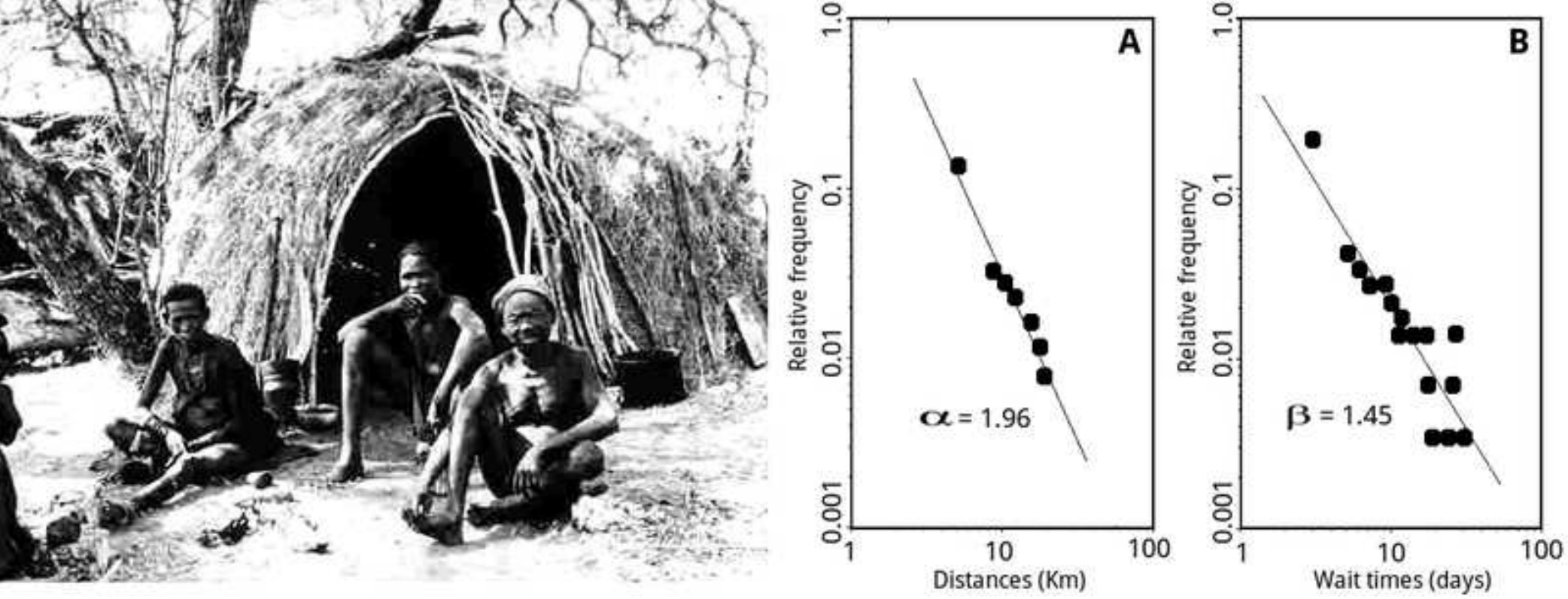}
\caption{{\small 
On human mobility and cooperation. No other aspect of the evolutionary biology is so impregnated by ideology as that of the nature of human cooperation. Fortunately scientific evidence is starting to prevail showing that mutual aid (reciprocity), as advanced by Kropotkin, is an extremely important factor that has shaped human evolution. Most of the evidence come from Game Theoretical results but most importantly from direct observation of human societies. Anthropological evidence has also provided great examples of human-environment interactions that clearly evince that human mobility patterns are landscape-driven\cite{miramontes2012non, boyer2012non} and that this may enhance social coherence and genetic diversity. One paradigmatic example was the hunter-gatherers San people of the southern Africa of the early twenty century (left picture) whose mobility patterns revealed power-law distributions in travelling distances\cite{brown2007levy} (A) and waiting times (B). These mobility patterns help explain why these human groups have the most genetic diversity of all the people on the planet\cite{tishkoff2009genetic}.       
}}
\label{san_people}
\end{figure}

\subsection{Stay together then interact}
Another front that must be included in a more comprehensive evolutionary theory of cooperation is the evolution of social interactions. It has been clear since the last two decades that social interactions obey a scale-free network pattern and that it seems to be ubiquitous in nature. Gene regulatory networks, metabolic networks, mutualistic networks, communications networks, etc. all of them seem to have long-tailed distributions, corresponding for instance to scale-free topologies. The reason for this is robustness and flux efficiency. Are scale-free networks a physical constraint in the origin and evolution of life? Can life-related networks, including social networks, have other topologies? One is tempted to answer no and the reason is simple. It is becoming apparent that an scale-free topology would facilitate the emergence of criticality in the dynamics running on them. This seems to suggest a bridge between the criticality of a phase transition and social dynamics. Examples of this are starting to emerge\cite{luque2008number}.

It was shown in models of ant-to-ant interactions, that a colony is posed at an order-disorder phase transition where sociality emerges and information capacity is at its best (Figure~\ref{insects}). In models of spider monkey foraging, it was found that the individual interactions with a given forest structure pose the ecosystem in a state where complex social networks emerge facilitated by the anomalous-diffusion nature of the animal displacements (Figures~\ref{monos} and \ref{monos-net}). A similar phenomenon has been recently revealed in ancient nomad human groups in Africa (Figure \ref{san_people}).

A novel theory of social evolution must integrate the concepts of the science of Complex Systems with those of the Darwinian tradition. Gene-centric concepts should be reviewed and complemented with evidence from multilevel phenomena (group selection), the constrains given by the non-linear nature of biological dynamical systems and the emergent nature of dissipative phenomena. Cooperation only emerges in come-and-stay-together scenarios, because of this, exploration of the properties of anomalous diffusion and the topological evolution of scale-free networks is very important. On the positive side, this research roadmap is on its way right now.

\subsection*{Acknowledgements}

We thank the late Dr. A. Chopps for many years of convivial and inspiring discussion, and Dr. Karo Michaelian,  Dr. Alessandra Marins and Dr. Paulo Cristaldo who kindly reviewed previous versions of this text providing invaluable insights. OM thanks DGAPA-PAPIIT Grant IN101712 and the Brazilian Ciência Sem Fronteiras (CSF-CAPES) program 0148/2012 for a Special Visiting Researcher position at the Universidade Federal de Viçosa, Brazil, during (2013-2015). ODS is supported by CNPq-Brasil, fellowship 305736/2013-2.


\bibliographystyle{IEEEtran-esp}
\bibliography{biblio}

\end{document}